\documentclass{PoS}

\title{Exclusive semileptonic $b \to c$ decays at Belle}

\ShortTitle{Exclusive semileptonic $b \to c$ decays at Belle}

\author{\speaker{Wolfgang DUNGEL}, \\%
        Institute for High Energy Physics, Austrian Academy of Sciences\\
        E-mail: \email{dungel@hephy.oeaw.ac.at}}

\author{on behalf of the Belle collaboration}
\abstract{We present analyses of exclusive semileptonic $b \to c$ decays based on data samples collected by the Belle detector at the KEK-B $e^+e^-$ asymmetric collider.

The first topic are precision measurements of the Cabibbo-Kobayashi-Maskawa matrix element $|V_{cb}|$ and the HQET form factor parameters $\rho^2$, $R_1$ and $R_2$ extracted from %$B^0 \to D^{*-} \ell+ \nu$ and 
$B^+ \to \bar{D}^{*0} \ell^+ \nu$ decays using untagged $\mathrm{\Upsilon}(4S)$ events. Additionally, a test of the form factor parametrization is performed.

Secondly, measurements of $B \to D^* \tau \nu$ and $B \to D \tau \nu$ decays are presented, where the accompanying second $B$ meson is tagged and reconstructed. Branching fractions of the semi-tauonic decays are measured.}

\FullConference{European Physical Society Europhysics Conference on High Energy Physics\\
		 July 16-22, 2009\\
		 Krakow, Poland}

\begin{document}

\section{KEKB and the Belle detector} % Section title should be in all capitals.

The following analyses were performed using data samples collected at the $\mathrm{\Upsilon}(4S)$ resonance with the Belle detector operating at the KEKB asymmetric-energy $e^+ e^-$ collider. The Belle detector is a large-solid-angle magnetic spectrometer that consists of a silicon vertex detector (SVD), a 50-layer central drift chamber (CDC), an array of aerogel Cherenkov counters (ACC), a barrel-like arrangement of time-of-flight scintillation counters (TOF) and an electromagnetic calorimeter comprised of CsI(Tl) crystals (ECL) located inside a superconducting solenoid coil that provides a 1.5 T magnetic field. An iron flux-return located outside the coil is instrumented to detect $K_L^0$ mesons and to identify muons (KLM). The detector is described in detail in Ref.~\cite{BelleDetector}.

\section{Exclusive $B^+ \rightarrow D^{*0} \ell^+ \nu$ decays}

From semileptonic decays of the kind $B \to D^{*} \ell \nu$ one can extract the Cabibbo-Kobayashi-Maskawa matrix
element~$|V_{cb}|$ times the form factor normalization $\mathcal{F}(1)$ as well as the HQET form factors $\rho^2$, $R_1$ and $R_2$. One of the dominating systematic error components is related with the slow pion emitted by the $D^*$ meson. Therefore it is worthwhile to investigate both $B^0$ and $B^+$ decays, where this systematic will not be identical.

\begin{figure}[hb]
  \begin{center}
    \includegraphics[width=0.35\columnwidth]{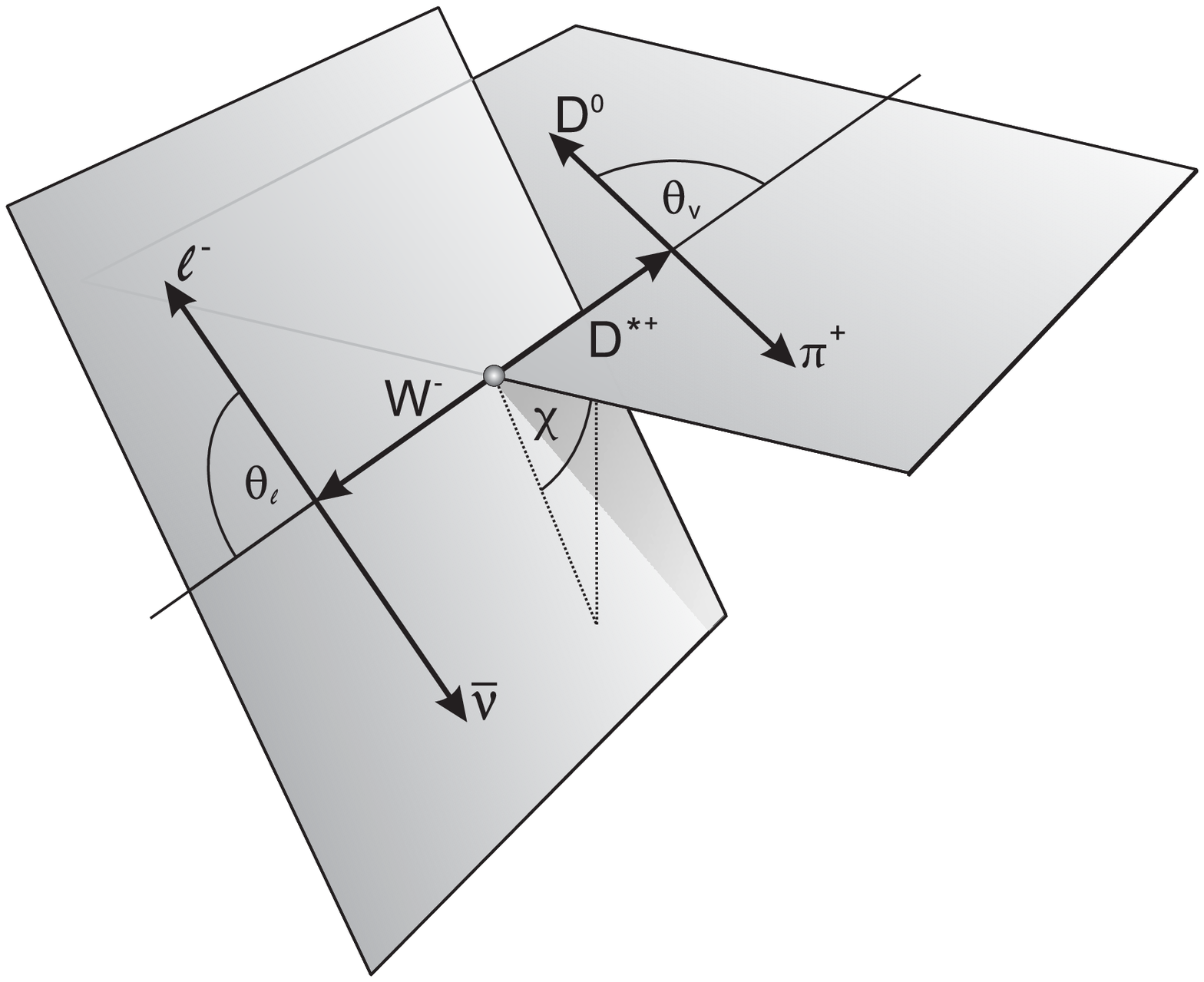}
    \hspace{0.05\columnwidth}
    \includegraphics[width=0.50\columnwidth]{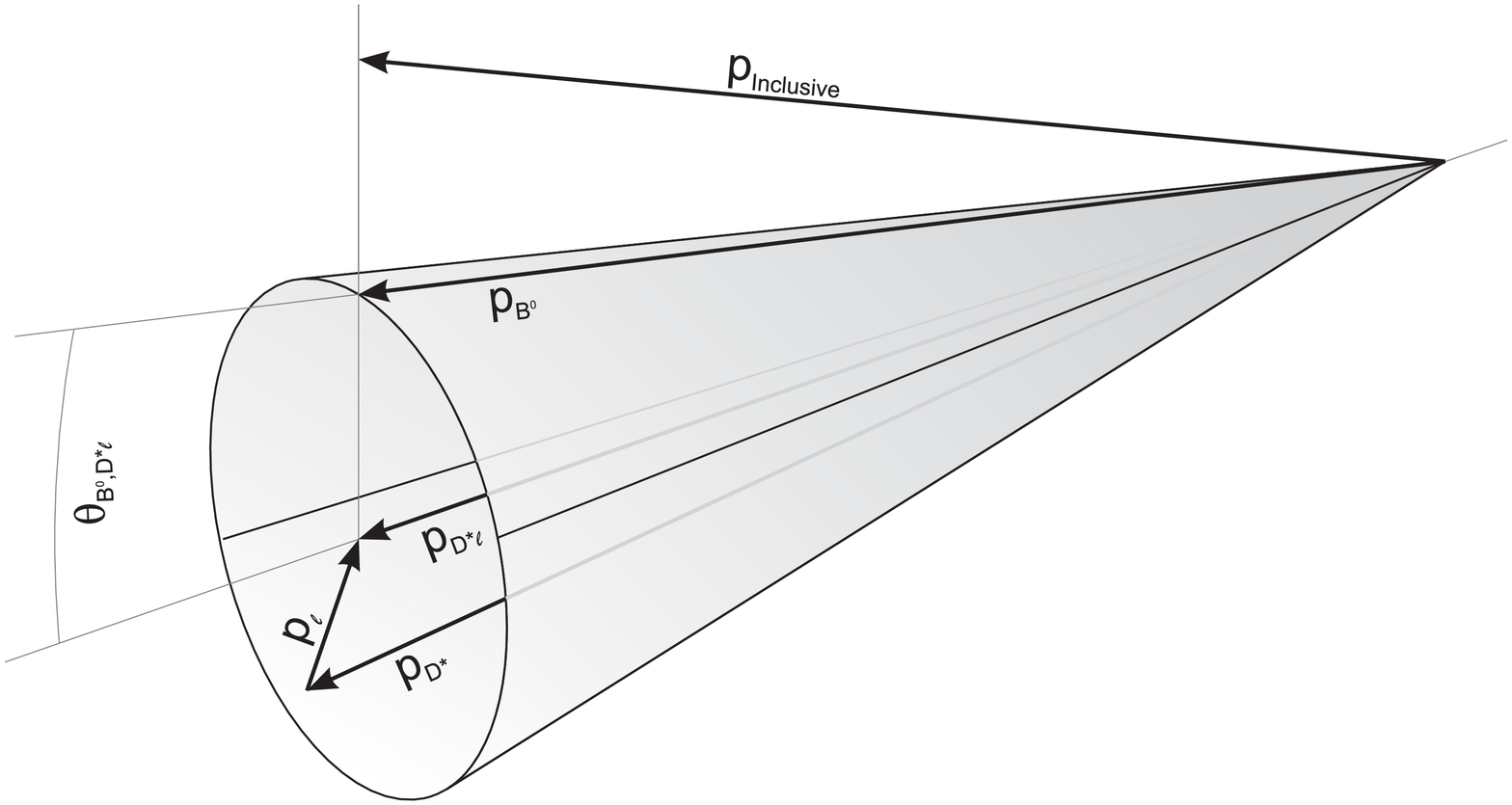}
  \end{center}
  \caption{The definition of the four kinematic variables $w $, $\cos \theta_{V}$, $\cos \theta_{\ell}$ and $\chi$ and a sketch of the reconstruction of the signal $B$ momentum using momentum conservation.}
  \label{fig:MeineKinematicVariablesReconstruction}
\end{figure}

The quadruple decay width of these processes is a four dimensional function depending on the kinematic variables~\cite{Neubert:1994,RichmanBurchat:1995} $w = v_{B}\cdot v_{D^{*}}$ and the three angles $\cos \theta_{V}$, $\cos \theta_{\ell}$ and $\chi$, defined in Fig.~\ref{fig:MeineKinematicVariablesReconstruction}. We use the parametrization defined in Ref.~\cite{CLN:1998}, which introduces the three free parameters $\rho^2$, $R_{1}(1)$ and $R_{2}(1)$ to govern the shape of the form factors of the decay.

Preliminary results based on the analysis of $B^0$ events have been reported before~\cite{Belle:Conf2008}. In the results presented here, the decay cascade $B^- \rightarrow D^{*0} \ell^- \bar{\nu}$, $D^{*0} \rightarrow D^{0} \pi^0_s$ with the slow pion $\pi^0_s \to \gamma\gamma$ and $D^{0} \rightarrow K^{-} \pi^+$ or $D^{0} \rightarrow K^{-} \pi^+ \pi^+ \pi^-$ is reconstructed. The light lepton $\ell$ is either an electron or a muon. Due to momentum conservation, the spatial momentum of the $B$ meson has to lie on a specific cone around the spatial momentum of the $D^{*}\ell$ system. The inclusive sum of the entire remaining event
%, which in case of perfect reconstruction yields the momentum of the second $B$ meson,
is used to obtain the best $B$ candiate by orthogonal projection, as sketched in the right hand plot of Fig.~\ref{fig:MeineKinematicVariablesReconstruction}. Data recorded about 60 MeV below the $\mathrm{\Upsilon}(4S)$ resonance is used to investigate background from $q\bar{q}$ decays, while Monte Carlo simulated events are used for a set of additional background components stemming from $B$ decays.

The parameters $\mathcal{F}(1) \left| V_{cb} \right|$, $\rho^2$, $R_{1}(1)$ and $R_{2}(1)$ are obtained by a binned least squares fit to the four one-dimensional marginal distributions of the decay width. The bin-to-bin correlations between these one dimensional histograms have to be considered. Only the branching ratio of the mode $D^{0} \rightarrow K^{-} \pi^+$ is used as an external parameter, the branching ratio of the mode $D^{0} \rightarrow K^{-} \pi^+ \pi^+ \pi^-$ is determined by fitting the ratio between the two $D^{0}$ channels, $R_{K3\pi/K\pi}$. A $\chi^2$ function is formed for each of the four channels separately and the sum of these four $\chi^2$'s is minimized numerically using the MINUIT package~\cite{James:1975dr}.

The preliminary results of the fit are $\rho^2=1.376\pm 0.074\pm 0.056$, $R_1(1)=1.620\pm 0.091\pm 0.092$,
$R_2(1)=0.805\pm 0.064\pm 0.036$, $R_{K3\pi/K\pi}= 2.072 \pm 0.023$, $\mathcal{B}( B^{+} \rightarrow \bar{D}^{*0} \ell^{+} \nu_{\ell} )= (4.84\pm 0.04\pm 0.56)\%$ and $\mathcal{F}(1)|V_{cb}|=35.0\pm 0.4\pm 2.2$, where the first error is the statistical error reported by MINUIT and the second (where shown) is the preliminary systematic error. The $\chi^{2} / \mathrm{n.d.f.}$ of the fit gives $187.8 / 155$.

%The results of the fit are shown graphically in Fig.~\ref{fig:MeineResultPlot}.
%
%\begin{figure}[h]
%  \begin{center}
%    \includegraphics[width=0.6\columnwidth]{fig/globalFit-sumOfChannels.eps}
%  \end{center}
%  \caption{Result of the fit of the four kinematic variables in the
%    total sample. (The different sub-samples are added in this plot.)
%    The points with error bars are continuum subtracted on-resonance
%    data. The histograms are the signal and the different background
%    components.}
%  \label{fig:MeineResultPlot}
%\end{figure}

\begin{figure}[h]
  \begin{center}
    \includegraphics[width=0.4\columnwidth]{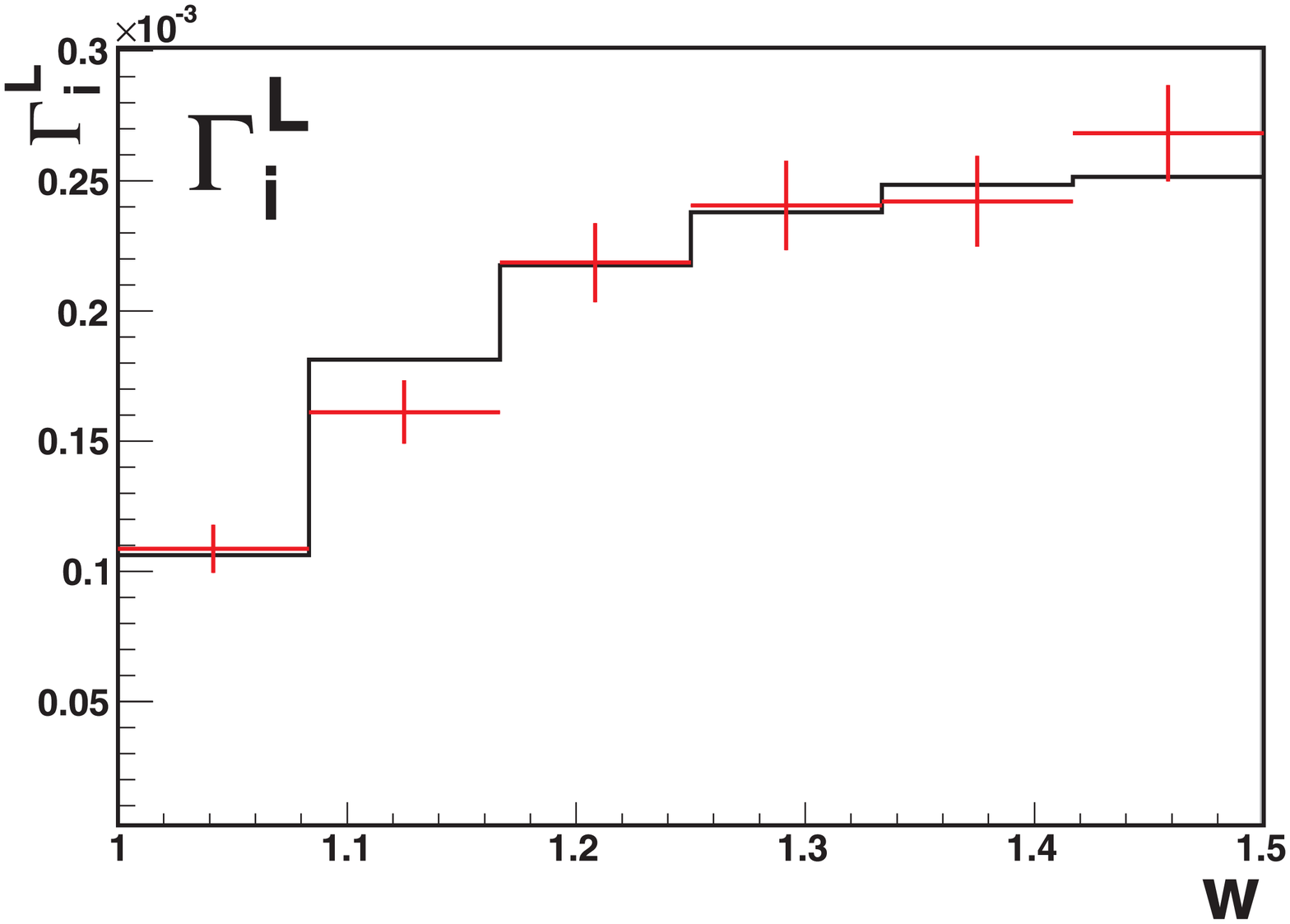}
    \includegraphics[width=0.4\columnwidth]{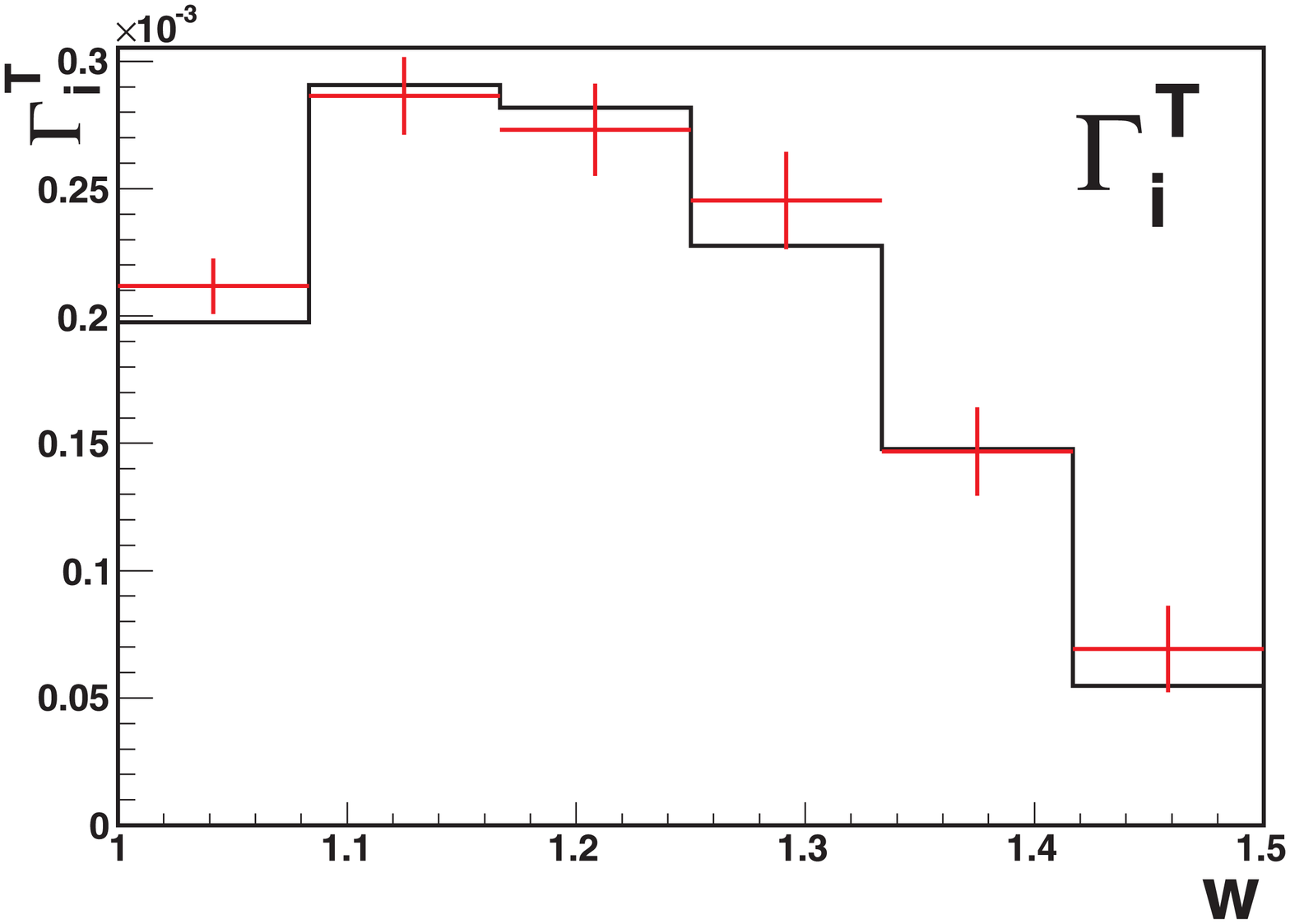}
  \end{center}
  \caption{Results of the fit of the helicity amplitudes (red crosses)
  compared to the prediction obtained by using the parameters obtained
  by using the parametrization prescription by Caprini, Lellouch and
  Neubert (solid black line). The left plot shows the results for
  $\Gamma^{L}_i$, the right one for $\Gamma^{T}_i$. Only the
  statistical error is shown. } \label{fig:8}
\end{figure}

Additionally, a cross check of the parametrization used to define the form factor parameters is performed by extracting the shapes of the longitudinal ($\Gamma^L$) and transversal ($\Gamma^T$) helicity amplitudes of the decay. There is good agreement between this cross check and the result by the parametrized fit, as shown in Fig.~\ref{fig:8}.

%      Part 1
%-----------------------------------------------------------------------------------
%      Part 2

\section{$B^0 \to D^{*-} \tau^+ \nu$ decays with inclusive tag of the accompanying $\bar{B}^0$}

Due to the large mass of the $\tau$ lepton, $b \to c \tau \nu$ decays can be used as probes for models containing charged Higgs bosons. The Belle collaboration was able to report the first observation of the decay $B^0 \to D^{*-} \tau^+ \nu$~\cite{KrakowBDstarTauNu}.

To reconstruct the signal, a $D^*$ meson is combined with a charged track, expected to stem from a $\tau$ decay. The $D^*$ candidates are reconstructed in the $D^* \to \bar{D}^0 \pi^-$ mode, with the $D^0$ being reconstructed in the channels $K\pi$ or $K\pi\pi^0$. The charged track associated with the $\tau$ decay is either an electron, $\tau^- \to e^- \bar{\nu}_e \nu_\tau$, or a charged pion, $\tau^- \to \pi^- \nu_\tau$. In case of the $\tau^- \to \pi^- \nu_\tau$ channel, only the $D^0 \to K\pi$ mode is used due to background constraints. All remaining tracks and clusters in the event are used to inclusively reconstruct the second B meson, $B_{\mathrm{tag}}$.

The variables $M_{\mathrm{tag}} = \sqrt{ E_{\mathrm{beam}}^2 - p_{\mathrm{tag}}^2 }$ and $\Delta E = E_{\mathrm{tag}} - E_{\mathrm{beam}}$ discriminate signal from background, since they can be used to check the consistency between the $B_{\mathrm{tag}}$ candidate and the $B$ hypothesis. Here $E_{\mathrm{beam}}$ is the beam energy and $E_{\mathrm{tag}}$ ($p_{\mathrm{tag}}$) is the energy (spatial momentum) of the sum of the residual particles. %Additional constraints are applied to improve the signal purity. These conditions are zero total event charge, no additional leptons in the event and zero total baryon number.

%Background suppression employs observables
%that are sensitive to missing energy in Bsig decay,
%like missing energy Emis = Ebeam . ED. . Ee,¥ð
%and visible energy of the event. The most effective
%variable Xmis, is defined by (Emis . |pD. +
%
%pe,¥ð|)/E2
%beam . m2
%B0 and is closely related to
%the missing mass in the Bsig decay.

\begin{figure}[h]
  \begin{center}
    \includegraphics[width=0.35\columnwidth]{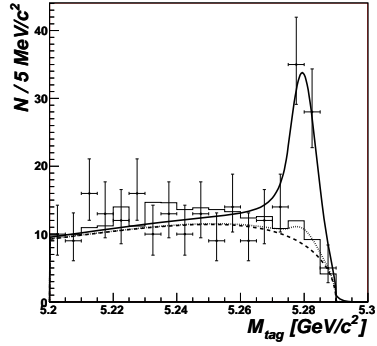}
  \end{center}
  \caption{$M_{\mathrm{tag}}$ distribution for $B^0 \to D^{*-} \tau^+ \nu$ decays. The real data is shown as points with error bars, the expected background is represented by the histogram. The solid line shows the fit result. The dotted and dashed curves indicate the distributions of background components.} 
  \label{fig:KrakowResult}
\end{figure}

The final signal yield is extracted by an unbinned maximum likelihood fit to the $M_{\mathrm{tag}}$ distribution, the result is shown in Fig.~\ref{fig:KrakowResult}. The fit result corresponds to $60^{+12}_{-11}$ events, which corresponds to $\mathcal{B}(B^0 \to D^{*-} \tau^+ \nu) = 2.02 ^{+0.40}_{-0.37}(stat) \pm 0.37 (syst) \% $. This value is consistent with Standard model expectations. Including systematic uncertainties, the significance of this result is $5.2 \sigma$.

%      Part 2
%-----------------------------------------------------------------------------------
%      Part 3

\section{Measurement of $B \to D^{(*)} \tau \nu$ using hadronic tag}

Measurements of the decays $B \to D \tau \nu$ suffer from large cross feeds from $B \to D^* \tau \nu$ decays. Therefore it is an efficient approach to measure both decays containing a $D$ and a $D^*$ meson within one analysis.

One $B$ meson is reconstructed in purely hadronic decay modes. For $B^+$ reconstruction, the modes $B^+ \to \bar{D}^{(*)0} \pi^+$, $B^+ \to \bar{D}^{(*)0} \rho^+$, $B^+ \to \bar{D}^{(*)0} a_1^+$ and $B^+ \to \bar{D}^{(*)0} D^{(*)+}_s$ are used, while $B^0$ candidates are obtained in the channels $B^0 \to D^{(*)-} \pi^+$, $B^0 \to D^{(*)-} \rho^+$, $B^0 \to D^{(*)-} a_1^+$ and $B^0 \to D^{(*)-} D^{(*)+}_s$. The selection of the $B_{tag}$ candidates is based on the beam-constraint mass $M_{\mathrm{bc}} = \sqrt{ E_{\mathrm{beam}}^2 - p_{\mathrm{tag}}^2 }$ and $\Delta E = E_{\mathrm{tag}} - E_{\mathrm{beam}}$, where $E_{\mathrm{tag}}$ ($p_{\mathrm{tag}}$) is the energy (spatial momentum) of the $B_{\mathrm{tag}}$ candidate in the c.m. system and $E_{\mathrm{beam}}$ is the beam energy. Events satisfying $5.27 < M_{\mathrm{bc}} < 5.29 \mathrm{GeV}/c^2$ and $-80 < \Delta E < 60 \mathrm{MeV}$ are considered to be well reconstructed $B$ events.

In the remaining event, a $D^{(*)}$ meson is reconstructed and a light lepton (electron or muon) emitted by the $\tau$ decay $\tau^- \to \ell^- \bar{\nu}_\ell \nu_\tau$ is looked for. $D$ candidates are reconstructed in the channels $\bar{D}^0 \to \{ K^+\pi^-,$ $K^+\pi^-\pi^0,$ $K^+ \pi^+\pi^-\pi^-,$ $K^+\pi^+\pi^-\pi^-\pi^0,$ $K^0_S \pi^0,$ $K^0_S \pi^+\pi^-,$ $K^0_S \pi^+\pi^- \pi^0\}$ and $D^- \to \{ K^+\pi^-\pi^-,$ $K^+\pi^-\pi^-\pi^0,$ $K^0_S \pi^-\}$. The invariant mass of the $D$ candidates has to be within $5\sigma$ of the nominal $D$ mass. The reconstructed $D^*$ modes are $D^{*+} \to D^0 \pi^+$, $D^{*+} \to D^+ \pi^0$, $D^{*0} \to D^0 \pi^0$ and $D^{*0} \to D^0 \gamma$. The mass difference $\Delta m = m_{D\pi(\gamma)} - m_D$ has to be within $5\sigma$ of the nominal value. The momentum of the lepton candidate, $p_\ell$, is required to be below $1.2 \mathrm{GeV}/c$. 

In signal events, one finds large values of  missing mass, defined by $M_{\mathrm{miss}}^2 = (E_{\mathrm{tag}} - E_D - E_\ell)^2 - (\vec{p}_{\mathrm{tag}} - \vec{p}_{D} - \vec{p}_\ell )^2$. The remaining, unmatched energy in the detector, $E^{\mathrm{ECL}}_{\mathrm{extra}}$, peaks at small values for signal. These two variables discriminate signal from background and a two-dimensional unbinned extended likelihood fit to these two distributions is used to extract the signal.

Due to large background contributions from $B\to D^{(*)} \ell^+ \nu$ decays, the branching ratios are not determined directly. Rather, the ratio $R(D^{(*)} \tau^+ \nu) = \mathcal{B}(B \to D^{(*)} \tau^+ \nu) / \mathcal{B}(B \to D^{(*)} \ell^+ \nu)$ is determined, which can then be multiplied with the world average of the $\ell$ modes to obtain the branching ratio of the $\tau$ modes. $R(D^{(*)} \tau^+ \nu)$ is suited for investigating possible impacts due to a charged Higgs, since it depends neither on the decay constant $f_B$ nor on the value of $\mathcal{B}(B \to D^{(*)} \ell^+ \nu)$.
The results are shown in Tab.~\ref{tab:4}.

\begin{table}
  \begin{center}
		\begin{tabular}{l|@{\extracolsep{.2cm}}ccc}
			\hline\hline
        Mode & $R= \mathcal{B}(B \to D^{(*)} \tau^+ \nu) / \mathcal{B}(B \to D^{(*)} \ell^+ \nu)$ & Statistical significance\\
        \hline
        $B^+ \to \bar{D}^0 \tau^+ \nu$		& $0.70^{+0.19}_{-0.18}(stat)^{+0.11}_{-0.09}(syst)$ & $3.8 \sigma$ \\
        $B^+ \to \bar{D}^{*0} \tau^+ \nu$	& $0.47^{+0.11}_{-0.10}(stat)^{+0.06}_{-0.07}(syst)$ & $3.9 \sigma$ \\
        $B^0 \to D^- \tau^+ \nu$					& $0.48^{+0.22}_{-0.19}(stat)^{+0.06}_{-0.05}(syst)$ & $2.6 \sigma$ \\
        $B^0 \to D^{*-} \tau^+ \nu$				& $0.48^{+0.14}_{-0.12}(stat)^{+0.06}_{-0.04}(syst)$ & $4.7 \sigma$ \\
 			\hline\hline
		\end{tabular}  
  \end{center}
  \caption{The measured ratios $R(D^{(*)} \tau^+ \nu)= \mathcal{B}(B \to D^{(*)} \tau^+ \nu) / \mathcal{B}(B \to D^{(*)} \ell^+ \nu)$ for each of the four signal modes and the statistical significance of the respective signal yields.} 
  \label{tab:4}
\end{table}

%      Part 3
%-----------------------------------------------------------------------------------
%      Extro
\section{Conclusions}

We presented three analyses based on data samples collected by the Belle detector at the KEKB $e^+e^-$ asymmetric energy collider: the determination of $\mathcal{F}(1) \left| V_{cb} \right|$ and the form factor parameters $\rho^2$, $R_1$ and $R_2$ in the exclusive decay $B^+ \rightarrow D^{*0} \ell^+ \nu$, the observation of the decay $B^0 \to D^{*-} \tau^+ \nu$ and the simultaneous measurement of the branching ratios of the four decays $B^+ \to \bar{D}^{(*)0} \tau^+ \nu$ and $B^0 \to D^{(*)-} \tau^+ \nu$.

%\begin{acknowledgments}
We thank the KEKB group for the excellent operation of the accelerator, the KEK
cryogenics group for the efficient operation of the solenoid, and the KEK computer group
and the National Institute of Informatics for valuable computing and Super-SINET network
support. 
%We acknowledge support from the Ministry of Education, Culture, Sports, Science,
%and Technology of Japan and the Japan Society for the Promotion of Science; the Australian
%Research Council and the Australian Department of Education, Science and Training; the
%National Natural Science Foundation of China under contract No. 10575109 and 10775142;
%the Department of Science and Technology of India; the BK21 program of the Ministry of
%Education of Korea, the CHEP SRC program and Basic Research program (grant No. R01-
%2005-000-10089-0) of the Korea Science and Engineering Foundation, and the Pure Basic
%Research Group program of the Korea Research Foundation; the Polish State Committee
%for Scientific Research; the Ministry of Education and Science of the Russian Federation and
%the Russian Federal Agency for Atomic Energy; the Slovenian Research Agency; the Swiss
%National Science Foundation; the National Science Council and the Ministry of Education
%of Taiwan; and the U.S. Department of Energy.
%\end{acknowledgments}

%--------------------------------------------------------------------------------------------------
%--------------------------------------------------------------------------------------------------

%--------------------------------------------------------------------------------------------------
%--------------------------------------------------------------------------------------------------

\end{document}